\documentclass[12pt,a4paper]{article}
\usepackage{amssymb,amsmath}
\usepackage{amsfonts}
\usepackage{a4}
\usepackage{enumerate,verbatim,tikz}

\newtheorem{theorem}{Theorem}
\newtheorem{lemma}{Lemma}

\begin{document}
 \baselineskip18pt
	
	\title{\bf  Skew Cyclic Codes Over A Finite Ring : A Note on a result of Mohammadi et al.}
	\author{Swati Bhardwaj{\footnote {E-mail: swatibhardwaj2296@gmail.com}} ~ and ~ Madhu Raka{\footnote {Corresponding author, e-mail: mraka@pu.ac.in}}
		\\ \small{\em Centre for Advanced Study in Mathematics}\\
		\small{\em Panjab University, Chandigarh-160014, INDIA}\\
		\date{}}
	\maketitle
\maketitle
{\abstract{ In this small note we correct an error made by Mohammadi et al. in their paper entitled ``On Skew Cyclic Codes Over A Finite Ring" ( Iranian Jl. Math. Sci. Inform. Vol 14 (1) (2019), 135-145). }}



\section{Introduction}

The concept of cyclic codes over $\mathbb{F}_{q}$ was generalized by Boucher et al. \cite{Bou2007} over a non-commutative ring, namely skew polynomial ring $\mathbb{F}_{q}[x;\theta]$, where $\mathbb{F}_{q}$ is a field with $q$ elements and $\theta$ is an automorphism of $\mathbb{F}_{q}$.
In the polynomial ring $\mathbb{F}_{q}[x;\theta]$, addition is defined as the usual one of polynomials and the multiplication is defined by the rule
$ax^i*bx^j=a\theta^i(b)x^{i+j}$ for $a,b \in \mathbb{F}_{q}$.  Many authors such as Boucher and Ulmer \cite{BU2009}, Siap et al. \cite{Siap2011}  studied skew cyclic codes over fields. Recently, several authors such as \cite{AM2018Asian}, \cite{BR}, \cite{Gur2014}, \cite{Kenza2019}, \cite{YSS2015},  studied skew cyclic codes over  finite rings.\vspace{2mm}

Let $S=\mathbb{F}_{p}+v\mathbb{F}_{p}+v^2\mathbb{F}_{p}$ be a finite ring, where $p$ is an odd prime and $v^3=v$.  Mohammadi et al. \cite{Moh2019} have discussed skew cyclic codes over $S$.
They proved in Proposition 2.4 that there is \textbf{only one} non-identity automorphism of $S$ given by  $\theta(a+bv+cv^2)=a-bv+cv^2$. If $p=2$, this is nothing but identity map, so it is necessary to take $p$ an odd prime. Mohammadi et al. \cite{Moh2019} did not specify it.\vspace{2mm}

\noindent \textbf{This result is not correct.}
 There are  six automorphisms of $S$. These are given by
\begin{equation}\label{eq1} \begin{array}{lll}
\theta_1(a+bv+cv^2)&=&a+b~v+c~v^2\vspace{2mm}\\
\theta_2(a+bv+cv^2)&=&a-b~v+c~v^2\vspace{2mm}\\
\theta_3(a+bv+cv^2)&=&(a+b+c)-\frac{b-c}{2}~v-\frac{3b+c}{2}~v^2\vspace{2mm}\\
\theta_4(a+bv+cv^2)&=&(a+b+c)+\frac{b-c}{2}~v-\frac{3b+c}{2}~v^2\vspace{2mm}\\
\theta_5(a+bv+cv^2)&=&(a-b+c)+\frac{b+c}{2}~v+\frac{3b-c}{2}~v^2\vspace{2mm}\\
\theta_6(a+bv+cv^2)&=&(a-b+c)-\frac{b+c}{2}~v+\frac{3b-c}{2}~v^2.
\end{array}\end{equation}

\noindent Here by $\frac{1}{2}$, we mean the inverse of $2$ in the multiplicative group $\mathbb{F}_{p}\setminus \{0\}$.\vspace{2mm}

\noindent While determining generator polynomials of skew cyclic codes over $S$, Mohammadi et al. \cite{Moh2019} strongly used that $\theta = \theta_2$.

\section{Automorphisms of $S$}
Let $p$ be an odd prime and $\theta$ be an automorphism of $S=\mathbb{F}_{p}+v\mathbb{F}_{p}+v^2\mathbb{F}_{p}$ $=\{a+bv+cv^2 : a,b,c \in \mathbb{F}_{p}\}$.  We know that the group of automorphisms of a finite field of characteristic $p$ is a cyclic group generated by the Frobenius map $a\mapsto a^p$.  So the restriction of $\theta$ over $\mathbb{F}_p$ is Identity automorphism. Therefore $\theta$ must look like $$\theta(a+bv+cv^2)=a+b\theta(v)+c\theta(v^2)=a+b\theta(v)+c(\theta(v))^2.$$ Hence to determine $\theta$, we need to evaluate $\theta(v)$.\vspace{2mm}

\begin{lemma}Let $z=a+bv+cv^2 \in S$. Then $z$ is a zero divisor if and only if either $a=0$ or $a-b+c=0$ or $a+b+c=0$.
\end{lemma}
\noindent This is Proposition 2.2 of \cite{Moh2019}.

\begin{theorem}
Let $p$ be an odd prime. If $\theta$ is an automorphism of $S=\mathbb{F}_{p}+v\mathbb{F}_{p}+v^2\mathbb{F}_{p}$ then,
\begin{equation}\begin{array}{ll}
\theta(v)=v &{\rm or,}\vspace{2mm}\\
\theta(v)=-v &{\rm or,}\vspace{2mm}\\
\theta(v)=1-\frac{1}{2}v-\frac{3}{2}v^2 &{\rm or,}\vspace{2mm}\\
\theta(v)=1+\frac{1}{2}v-\frac{3}{2}v^2 &{\rm or,}\vspace{2mm}\\
\theta(v)=-1+\frac{1}{2}v+\frac{3}{2}v^2 &{\rm or,}\vspace{2mm}\\
\theta(v)=-1-\frac{1}{2}v+\frac{3}{2}v^2.
\end{array}\end{equation}
\end{theorem}

\noindent \textbf{Proof:}
Let $\theta$ be an automorphism of $S=\mathbb{F}_{p}+v\mathbb{F}_{p}+v^2\mathbb{F}_{p}$. Let $\theta(v)=x+yv+zv^2$. Then
\begin{equation*} \begin{array}{ll}
x+yv+zv^2&= \theta(v) =\theta(v^3) =(x+yv+zv^2)^3\vspace{2mm}\\
&=x^3+(3yz^2+6xyz+y^3+3x^2y)v\vspace{2mm}\\
&+(z^3+3xz^2+3y^2z+3x^2z+3xy^2)v^2.
\end{array}\end{equation*}
Therefore,
\begin{equation}\label{eq2} \begin{array}{ll}
&x=x^3\\
&y=3yz^2+6xyz+y^3+3x^2y\\
&z=z^3+3xz^2+3y^2z+3x^2z+3xy^2.
\end{array}\end{equation}

\noindent Hence $x$ can be $0,1$ or $-1$. We distinguish these in the following cases. \vspace{2mm}

\noindent{\textbf{Case I :}} $x=0$ \vspace{2mm}\\
From (\ref{eq2}), we get
\begin{equation}\label{eqn3} \begin{array}{ll}
3yz^2+y^3&= y \\
z^3+3y^2z&= z.
\end{array}\end{equation}

\noindent \textbf{Subcase (i) :} $y=0$ \vspace{1mm}\\
From (\ref{eqn3}), we have $z^3=z$ i.e. $z=0$ or $\pm 1$.\\
Clearly $z \neq 0$, otherwise it will be a zero map which is not an automorphism.\\
Let $z=1$ then $\theta(v)=v^2$. As $\theta$ is an automorphism, $\theta(v^2)=(\theta(v))^2=v^4=v^2=\theta(v)$ as $v^3=v$, implying that $\theta$ is not one-one. If $z=-1$ we get, $\theta(v)=-v^2$, therefore $\theta(v^2)=(\theta(v))^2=(-v^2)^2= v^4=v^2=\theta(-v)$, which again prevents $\theta$ to be one-one.\vspace{2mm}

\noindent \textbf{Subcase (ii) :} $y \neq 0$ \vspace{1mm}\\
Here from (\ref{eqn3}), we have $3z^2+y^2=1$ and $z(z^2+3y^2-1)=0$.\\
If $z=0$, we have $y^2=1$, i.e. $y=\pm 1$.\vspace{1mm}\\
If $y=1$ then $\theta$ is nothing but identity automorphism, and if $y=-1$ then $\theta(v)=-v$.\\
Now let $z \neq 0$ then $z^2+3y^2-1=0$ and $3z^2+y^2-1=0$. On solving these, we get $z^2=1/4$ and $y^2=1/4$, i.e. $z=\pm 1/2$ and $y=\pm 1/2$. In all these four possibilities, we will show that $\theta(v)=\theta(v^2)$. Suppose  $y=-1/2$ and $z=1/2$. Then $\theta(v^2)=(\theta(v))^2=(-v/2+v^2/2)^2=v^2/4+v^4/4-v^3/2=-v/2+v^2/2=\theta(v)$ as $v^3=v$. Similarly in other three possibilities we get that $\theta(v)=\theta(v^2)$, implying that $\theta$ is not one-one.  \vspace{2mm}

\noindent \textbf{Case II :} $x=1$ \vspace{2mm}\\
Since $v$ is zero divisor and $\theta$ is an automorphism, $\theta(v)=1+yv+zv^2$ is also a zero divisor. Therefore, by Lemma 1, either $1+y+z=0$ or $1-y+z=0$,  i.e. $y=\pm (z+1)$.\vspace{2mm}

\noindent From (\ref{eq2}), we have
\begin{equation}\label{eq3}
 3yz^2+6yz+y^3+3y=y.
 \end{equation}

 \noindent Substituting $y=\pm (z+1)$ in (\ref{eq3}) we get
$$ 4z^3+12z^2+11z+3=0$$ which implies
 $$ (z+1)(2z+1)(2z+3)=0, ~{\rm i.~e.}~z=-1{\rm ~or~} \frac{-1}{2}{\rm ~or~} \frac{-3}{2}.$$

 \noindent \textbf{Subcase (i) :} $y=z+1$ \vspace{1mm}\\
 If $z=-1$ then $y=0.$ This gives $\theta(v)=1-v^2$, in which case $\theta(v)=\theta(v^2) $ and hence $\theta$ is not one-one. If $z=-1/2$ then $y=1/2$, so $\theta(v)=1+v/2-v^2/2$. Here again, $\theta(v^2)=(\theta(v))^2=(1+v/2-v^2/2)^2=1+v/2-v^2/2=\theta(v)$ as $v^3=v$, a contradiction.\\
 If $z=-3/2$ then $y=-1/2$ and so $\theta(v)=1-v/2-3v^2/2.$  \vspace{2mm}

 \noindent \textbf{Subcase (ii) :} $y=-(z+1)$ \vspace{2mm}

\noindent Working as in subcase (i), one can see that the only possibility for $\theta$ to be an automorphism is when
$\theta(v)=1+v/2-3v^2/2$.\vspace{2mm}

\noindent \textbf{Case III :} $x=-1$ \vspace{2mm}

 \noindent We have from (\ref{eq2})
 \begin{equation}\label{eq5}
 3yz^2-6yz+y^3+3y=y.
 \end{equation}

\noindent As $\theta(v)=-1+yv+zv^2$ is a zero divisor, by Lemma 1 we must have either $-1+y+z=0$ or $-1-y+z=0$.\\
Substituting $y=\pm (z-1)$ in (\ref{eq5}) we find that
$$4z^3-12z^2+11z-3=0, {\rm ~ i.e.,} ~(z-1)(2z-1)(2z-3)=0.$$
Working as in Case II, we find that the only possibilities for $\theta$ to be an automorphism are $\theta(v)=-1+v/2+3v^2/2$ and $\theta(v)=-1-v/2+3v^2/2$.

\begin{theorem}
There are exactly six automorphisms of $S$.
\end{theorem}

\noindent \textbf{Proof:}
We will show that each of six values of $\theta(v)$ obtained in Theorem 1 lead to six distinct automorphisms of $S$ listed in (\ref{eq1}). If $\theta(v)=v$, then  $\theta$  is identity on $S$. If $\theta(v)=-v$, then $\theta(v^2)=(\theta(v))^2=v^2$. Therefore $\theta(a+bv+cv^2)=a-bv+cv^2=\theta_2(a+bv+cv^2)$. Clearly $\theta_1$ and $\theta_2$ are one-one and onto. \vspace{2mm}

\noindent Suppose $\theta(v)=1-\frac{1}{2}v-\frac{3}{2}v^2$.  In this case, $\theta(v^2)=(\theta(v))^2=(1-\frac{1}{2}v-\frac{3}{2}v^2)^2=1+\frac{1}{2}v-\frac{1}{2}v^2$, and also $\theta(1)=1$. Thus $$\begin{array}{ll}\theta(a+bv+cv^2)&=a+b(1-\frac{1}{2}v-\frac{3}{2}v^2)+c(1+\frac{1}{2}v-\frac{1}{2}v^2)
\vspace{2mm}\\&=(a+b+c)-\left(\frac{b-c}{2}\right)v-\left(\frac{3b+c}{2}\right)v^2\vspace{2mm}\\&=\theta_3(a+bv+cv^2).\end{array}$$
One finds that $\theta_3$ is one-one and onto. This is so
because if $\theta_3(a+bv+cv^2)=0$, we get $a+b+c=0,~\frac{b-c}{2}=0,{\rm ~and~}~ \frac{3b+c}{2}=0$, which gives $a=b=c=0$.
And, for any $x+yv+zv^2 \in S,$ we have $\theta_3\big((x-y+z) - (\frac{y+z}{2})v + (\frac{3y-z}{2})v^2\big) = x+yv+zv^2$, showing thereby that $\theta_3$ is onto.  Thus, $\theta_3$ is an automorphism.\\
Similarly, one can easily check that all the remaining  values of $\theta(v)$ obtained in Theorem 1 lead to  automorphisms of $S$.



\begin{thebibliography}{99}
\bibitem{AM2018Asian} M. Ashraf, G. Mohammad, Skew-cyclic codes over $\mathbb{F}_{q}+u\mathbb{F}_{q}+v\mathbb{F}_{q}$, \emph{Asian-European Journal of Mathematics}, \textbf{11} (5), (2018), 1850072, doi.org/10.1142/S1793557118500729. \vspace{1mm}

\bibitem{BR} S. Bhardwaj, M. Raka, Skew constacyclic codes over a non-chain ring $\mathbb{F}_{q}[u,v]/\langle f(u),g(v), uv-vu\rangle$, \emph{Appl. Algebra Engg. Comm. Comput.} \textbf{31} (3), (2020), 173-194, doi.org/10.1007/s00200-020-00425-z.\vspace{1mm}

\bibitem {Bou2007} D. Boucher, W. Geiselmann, F. Ulmer, Skew cyclic codes, \emph{Appl. Algebra Engg. Comm. Comput.}, \textbf{18} (4), (2007), 379-389.\vspace{1mm}

\bibitem {BU2009} D. Boucher, F. Ulmer, Coding with skew polynomial ring, \emph{J. Symb. Comput.} \textbf{44} (12), (2009), 1644-1656. \vspace{1mm}

\bibitem{Gur2014} F. Gursoy, I. Siap, B. Yildiz, Construction of skew cyclic codes over $\mathbb{F}_q+v\mathbb{F}_q$, \emph{Adv. Math. Commun.}, \textbf{8} (3), (2014), 313-322 .\vspace{1mm}

 \bibitem{Kenza2019} J. Kabore, A. Foutue-Tabue, K. Guenda, M.E. Charkani, Skew-constacyclic codes over $\mathbb{F}_q[v]/\langle v^q-v \rangle$, arXiv:1902.10477v1[cs.IT], (2019).\vspace{1mm}

\bibitem{Moh2019} R. Mohammadi, S. Rahimi, H. Mousavi, On skew cyclic codes over a finite ring, \emph{Iranian J. Math. Sci. Inform.}, \textbf{14}(1), (2019), 135-145 .\vspace{1mm}

\bibitem{Siap2011} I. Siap, T. Abualrub, N. Aydin, P. Seneviratne, Skew cyclic codes of arbitary length, \emph{Int. J. Inform.  Coding Theory},  \textbf{2} (1), (2011), 10 -20.  \vspace{1mm}		 
			
\bibitem{YSS2015} T. Yao, M. Shi, P. Sol$\acute{e}$, Skew cyclic codes over $\mathbb{F}_{q}+u\mathbb{F}_{q}+v\mathbb{F}_{q}+uv\mathbb{F}_{q}$, \emph{J. Algebra Comb. Discrete Appl. }, \textbf{2}(3), (2015), 163-168. 	


\end{thebibliography}
\end{document}